\documentclass{aa} 
\usepackage{epsfig,times} 

\def\mincir{\raise -2.truept\hbox{\rlap{\hbox{$\sim$}}\raise5.truept \hbox{$<$}\ }} 
\def\magcir{\raise -2.truept\hbox{\rlap{\hbox{$\sim$}}\raise5.truept \hbox{$>$}\ }} 
\def\mincireq {\hbox{\raise0.5ex\hbox{$<\lower1.06ex\hbox{$\kern-1.07em{\sim}$}$}}} 
\def\magcireq {\hbox{\raise0.5ex\hbox{$>\lower1.06ex\hbox{$\kern-1.07em{\sim}$}$}}} 
\def\gr{\kern 2pt\hbox{}^\circ{\kern -2pt K}} 

\def\rosat{{\sl ROSAT }}
\def\asca{{\sl ASCA }}

\def\chandra{{\sl Chandra }}

\def\a{\alpha}
\def\b{\beta}

\def\G{\Gamma}

\def\s{\sigma}

\def\_{\thinspace}

\def\cm{\thinspace{\rm cm}}
\def\cm2{\thinspace{\rm cm}^2}

\def\be{\begin{equation}}
\def\ee{\end{equation}}

\begin{document} 

\titlerunning{X-rays from NGC~4666}
\title{The composite starburst/AGN nature of the superwind galaxy NGC~4666}

\author{M. Persic\inst{1}, 
	M. Cappi\inst{2},
	Y. Rephaeli\inst{3,4},
	L. Bassani\inst{2},
	R. Della Ceca\inst{5},
	A. Franceschini\inst{6},
        L. Hunt\inst{7},
	G. Malaguti\inst{2},
and
	E. Palazzi\inst{2}} 
\institute{ 
INAF/Osservatorio Astronomico di Trieste, via G.B.Tiepolo 11, I-34131 
Trieste, Italy
	\and
IASF/CNR -- Sezione di Bologna, via P.Gobetti 101, I-40129 Bologna, Italy
	\and  
School of Physics and Astronomy, Tel Aviv University, Tel Aviv 69978, Israel
	\and
CASS, University of California, San Diego, La Jolla, CA 92093, USA
	\and
INAF/Osservatorio Astronomico di Brera, via Brera 28, I-20121 Milano, Italy
	\and
Dipartimento di Astronomia, Universit\`a di Padova, vicolo Osservatorio 2, I-35122 Padova, Italy 
	\and
IASF/CNR -- Sezione di Firenze, l.go E.Fermi 5, I-50125 Firenze, Italy }

\date{Received ..................; accepted ...................} 

\offprints{M.P.; e-mail: {\tt persic@ts.astro.it}}

\abstract{ 

We report the discovery of a Compton-thick 
AGN and of intense star-formation activity in the nucleus and disk, respectively, of the 
nearly edge-on superwind galaxy NGC~4666. Spatially unresolved emission is detected by 
{\it BeppoSAX} only at energies $<10$ keV, whereas spatially resolved emission from the whole disk 
is detected by {\it XMM-Newton}. A prominent (EW $\sim 1-2$ keV) emission line at $\sim$6.4 keV is 
detected by both instruments. From the {\it XMM-Newton} data alone the line is spectrally localized 
at $E \simeq 6.42 \pm 0.03$ keV, and seems to be spatially concentrated in the nuclear 
region of NGC~4666. This, together with the presence of a flat ($\G \sim 1.3$) continuum 
in the nuclear region, suggests the existence of a strongly absorbed (i.e., Compton-thick) 
AGN, whose intrinsic 2-10 keV luminosity is estimated to be $L_{2-10} \magcir 2 \times 
10^{41}$ erg s$^{-1}$. 
At energies $\mincir$1 keV the integrated ({\it BeppoSAX}) spectrum is dominated by a $\sim$0.25 
keV thermal gas component distributed throughout the disk (resolved by {\it XMM-Newton}). At energies 
$\sim$2-10 keV, the integrated spectrum is dominated by a steep ($\G \magcir 2$) power-law 
(PL) component. The latter emission is likely due to unresolved sources with luminosity 
$L \sim 10^{38} - 10^{39}$ erg s$^{-1}$ that are most likely accreting binaries (with 
BH masses $\leq$8$M_\odot$). Such binaries, which are known to dominate the X-ray 
point-source luminosity in nearby star-forming galaxies, have $\G \sim 2$ PL spectra in 
the relevant energy range. A $\G \sim 1.8$ PL contribution from Compton scattering of 
(the radio-emitting) relativistic electrons by the ambient FIR photons may add a truly 
diffuse component to the 2-10 keV emission. 

}

\maketitle
\titlerunning
\authorrunning
\markboth{M.Persic et al.: X-rays from NGC~4666}{} 
 
\section{Introduction} 

The Compton-thick subclass of Seyfert 2 (Sy2) galaxies consists of objects 
for which the column density of absorbing matter along the l.o.s. to 
the circumnuclear torus is $N_{\rm H} > \sigma_T^{-1}  \simeq 1.5 
\times 10^{24}$ cm$^{-2}$. A column of a few times this critical value 
is sufficient to strongly depress the transmitted intensity also in 
hard X-rays, since after a few scatterings the radiation is redshifted 
into the photoelectrically-dominated regime. 

However, even for very high values of $N_{\rm H}$, nuclear radiation 
can be observed by the radiation scattered from {\it (a)} the visible 
part of the inner surface of the circum-nuclear torus (that blocks 
direct X-ray emission entirely from our view; Antonucci 1993), and from 
scattering by {\it (b)} the warm medium that scatters and polarizes 
the optical broad lines (Antonucci \& Miller 1985). The two scattered 
components are spectrally different.
\smallskip

\noindent
{\it (a)} The torus, which is presumed to be essentialy neutral, gives rise 
to the 'Compton reflection continuum' or 'cold reflection' (see George \& 
Fabian 1991) that can be explained as follows. 
Suppose an input power-law (PL) continuum irradiates some 'cold' (i.e., 
$T \mincir 10^6 \gr$) gas with solar chemical abundance. The reflected X-ray 
spectrum contains the Fe-K$\a$ (6.40 keV), Fe-K$\b$ (7.06 keV), Ni-K$\a$ 
(7.48 keV) fluorescence lines, along with the Fe-K-shell absorption edge 
at $\sim$7.11 keV. Due to the high abundance and fluorescence yield of 
iron, a significant fraction of the incident radiation at energies $E \geq 
7.11$ keV yields Fe-K$\a$ fluorescent line at $E = 6.40$ keV. Compton 
down-scattering within the gas causes a fall-off in the continuum at high 
energies, a smearing of the Fe-K-shell absorption edge, and the formation of 
low-energy wings of the fluorescence lines. The competition within the slab 
between multiple electron down-scattering of high energy photons and 
photoelectric absorption of low-energy photons leads to a 'Compton reflection 
hump' in the reflected X-ray continuum in the $\sim$ 20-100 keV band.
\smallskip

\noindent
{\it (b)} The warm mirror, which is presumed to be highly ionized  (i.e., 
$T \magcir 10^7 \gr$) and of low optical thickness, is located above the 
obscuring torus and electron-scatters the continuum photons coming from 
the central source (Krolik \& Kallman 1987). If the ionized scattering 
medium is optically thin, photoelectric absorption is of little importance 
on the electron-scattered continuum, so the emergent spectrum essentially 
retains the spectral shape of the primary emission (except for a high-energy 
cutoff due to Compton down-scattering). This same warm scattering medium 
most likely produces the higher-energy Fe-K$\a$ lines from, respectively, 
He-like iron (Fe~XXV) at $\sim$6.7 and H-like iron (Fe~XXVI) at 6.97 keV. 

A combination of warm and cold reflections is, therefore, a plausible 
model for the observed X-ray spectra of Sy2 sources hidden in galactic 
nuclei (Matt et al. 1996, 1997). The relative strength of the cold and 
warm reflections can possibly be determined from fitting the composite 
model to 2-10 keV \asca measurements of the proto-typical Sy2 galaxy 
NGC~1068. From the fraction of 2-10 keV flux that comes from cold 
reflection, and an independent estimate of the intrinsic $L_{2-10}^\prime 
\simeq 7 \times 10^{43}$ erg s$^{-1}$ from the [O~III]$\lambda$5007 
luminosity, and a 2-10 keV albedo of 0.022 for the whole inner wall of 
the torus, Iwasawa et al. (1997) deduced that a fraction $f_{\rm C} 
\simeq 4.4 \times 10^{-2} (L^\prime_{43.8}) ^{-1}$ of the inner surface 
of the torus was visible to us. Such a small fraction of the reflecting 
inner surface of the torus indicates that the torus is viewed nearly 
edge-on. Iwasawa et al. (1997) further deduced the warm-scattered luminosity 
fraction to be $f_{\rm w} \simeq 7.8 \times 10^{-4} (L^\prime_{43.8})^{-1}$. 
The warm-to-cold reflection luminosity ratio in NGC~1068 is then 0.018. 
For a less extreme inclination of the absorbing torus -- i.e. for a 
higher fraction of the reflecting surface -- the 2-10 keV spectrum of a 
Sy2-like source is even more dominated by cold reflection. Thus, we may 
conclude that the spectral features of cold reflection, i.e. a 'flat' 
($\G \sim 1.3$) 2-10 keV continuum and a 6.4 keV K$\a$ line from 'cold' 
iron, constitute the X-ray signature of a strongly absorbed (i.e., 
Compton-thick) Sy2-like source. Based on these results, a strategy aimed 
at unveiling strongly absorbed Active Galactic Nuclei (AGNs) 
harbored in galactic nuclei is to make spectral observations of galaxies in 
the band 2-10 keV, and to look for a prominent 6.4 keV line superimposed on 
a rather flat underlying continuum.

A substantial link between circum-nuclear starbursts (SBs) and AGNs may be 
established if both occur in the same galaxy. If the SB-driven turbulence in 
the gas increases the mass inflow rate onto a central black hole, a higher 
emission from the central engine will ensue (e.g., Veilleux 2001). Since a 
circum-nuclear SB is located farther out than the torus and its obscuring 
material is distributed quasi-isotropically as seen from the central source, 
the nuclear source will be further obscured by the SB so that the emerging 
(cold-reflected) X-ray spectrum will be further hardened by SB-related 
photoelectric absorption (Fabian et al. 1998). Roughly 50$\%$ of all Sy2 
galaxies contain circumnuclear SBs (Gonzales Delgado et al. 2001): these 
composite objects do tend to be more obscured than those lacking SBs. It is 
indeed reasonable to conjecture that such extra obscuration is caused by the 
SB, because star-formation rates (SFRs) per unit area of $\sim$50-100 
$M_\odot$ yr$^{-1}$ kpc$^{-2}$ (typical of the innermost 0.1 kpc in such 
composite galaxies; see Gonzales Delgado et al. 1998) imply $N_{\rm H} \sim 
10^{24}$ cm$^{-2}$ (see Kennicutt 1998). A circum-nuclear SB can also be 
physically connected to a torus. If the mass inflow rate through the accretion 
disk is higher than the rate of power dissipation in the disk, its opacity will 
increase, and the disk will puff up into a geometrically thick torus with a 
large covering angle (as seen from the central engine). This will quickly block 
the primary AGN continuum and will result in emission of the fluorescent Fe-K$\a$ 
line and flat reflection continuum. Indeed, among the heavily obscured ($N_{\rm H} 
> 10^{24}$ cm$^{-2}$) sources studied by Levenson et al. (2002), those that exhibit 
the largest equivalent widths (EWs) have concentrated circumnuclear SBs, 
whereas in the only galaxy certainly lacking a SB, Mrk 3, the Fe-K$\a$ line 
is relatively weak (EW $\sim 0.7$).

Well-known examples of coexisting spectral signatures of star formation and 
Compton-thick AGNs include cases of both type-1 and type-2 view of the central 
AGN. (These views define Sy1 and Sy2 morphology in the `Unified AGN 
Model', e.g. Urry \& Padovani 1995.) Examples of the former type include Arp~299, 
NGC~6240, and NGC~4945: in these objects the absorbing matter is moderately thick 
to Compton scattering ($N_{\rm H} \sim$ a few $\times 10^{24}$ cm$^{-2}$), so the 
the primary AGN spectrum emerges at (source restframe) energies $E \magcir 10$ 
keV, within the {\it BeppoSAX} observing window (Della Ceca et al. 2002 and Ballo et al. 
2004; Vignati et al. 1999; Guainazzi et al. 2000). The archetypal example of the 
latter type is NGC~1068 for which {\it BeppoSAX} data support a model envisaging a mixture 
of both cold and warm reflections of an otherwise unseen primary nuclear continuum, 
the cold reflection component being the dominant one in the 20-100 keV band (Matt 
et al. 1997). 

NGC~4666 is a starburst galaxy (SBG) seen nearly edge-on 
(see Table 1). Based on multi-frequency observations, Dahlem et al. (1997) 
have detected an outflow cone -- associated with a galactic superwind -- 
emanating from a central SB of $\sim$3.3 kpc in radius, and having an opening 
angle of $30^\circ \pm 10^\circ$. The outflow is traced up to $\sim$7.5 kpc 
above the disk plane by optical emission line filaments, nonthermal radio 
continuum emissiom, and soft X-ray emission from hot gas. The multi-wavelength 
evidence for enhanced star-formation activity in NGC~4666 can be summarized as 
follows:

\noindent
{\it (i)} 
Optical spectroscopy exhibits the kinematic signature of an outflow, and the 
observed line ratios indicate shock-heating [e.g., following SB-related supernova 
{(SN)} explosions] as the most likely excitation mechanism for the emission-line 
gas.

\noindent
{\it (ii)} 
The existence of a synchrotron radio halo (Sukumar et al. 1988; Dahlem et al. 1997) 
is strong evidence of ongoing or very recent SB activity. The radio (energy) index 
ranges from $\a_r \sim 0.8$, measured at small galactocentric radii, to  $\a \magcir 
1.3$, measured at large radii. This is proof of the presence, close to the disk, of 
ongoing injection of fresh relativistic electrons whose lifetime against radiative 
losses, $t = 10^{10} (E/{\rm GeV})^{-1} (B/\mu{\rm G})^{-2}$ yr, is $t \sim 5-50$ Myr 
[for $B \sim 15\, \mu$G (Dahlem et al. 1997) and $E \sim 1-10$ GeV], i.e. comparable 
or shorter than a typical SB age ($t_{\rm SB} \sim 10^8$ yr). The degree of steepening 
is just what is expected when the electrons lose energy radiatively. 

\noindent
{\it (iii)} 
At energies $<$2.4 keV NGC~4666 was previously observed with \rosat (Dahlem 
et al. 1997, 1998); the $4^{\prime\prime}$-resolution HRI image showed faint 
diffuse emission extending along the major axis and confined mostly to the 
disk plane, and two point sources located $\sim 1^\prime$ along the minor 
axes above and below the nuclear region. The spatially integrated PSPC spectrum 
was fit with a two-component model (Dahlem et al. 1998) consisting of a PL + 
thermal model, with the PL photon index set at $\G=1.9$, and a derived plasma 
temperature (with the chemical abundances set at the solar value) of $kT=0.31^
{+0.31}_{-0.12}$ keV. 

\begin{table*}
\caption[] { General characteristics of NGC~4666.}
\begin{flushleft}
\begin{tabular}{ l  rl  l  l l  l  l l  l   l l l }
\noalign{\smallskip}
\hline
\noalign{\smallskip}
Type$^{(a)}$ & RA$^{(b)}$  & DEC$^{(b)}$ & D$^{(c)}$ & $B_{\rm T}^{\rm i,0 \,(d)}$ & log$L_{\rm B}^{0\,(e)}$ &
  log$N_{\rm H}^{(f)}$ & log$L_{\rm FIR}^{(g)}$ & incl$^{(h)}$ & $R_{25}^{{\rm i,0} \,(i)}$ & 
log$L_{0.5-2}^{(j)}$ & log$L_{2-10}^{(k)}$ \\
        &(J2000) &(J2000) &  (Mpc)     &   (mag)  & ($L_{\odot}$) &
  (cm$^{-2}$) & ($L_{\odot}$) & (deg) & (kpc) & (erg/s) & (erg/s)  \\
\noalign{\smallskip}
\hline
\noalign{\smallskip}
 SABc & $12^h 45^m 09^s$ & $-00^{\circ} 27^{\prime} 38^{\prime\prime}$ & 26.3 &
 10.68 & 10.76  &  20.23  &  10.69 & $80^{\circ}$ & 17.5 & 40.20 & 40.40 \\
\noalign{\smallskip}
\hline
\end{tabular}
\end{flushleft}
\bigskip

$^{(a)}$ Morphological type, from {\sl RC3} (de Vaucouleurs et al. 1991).  

$^{(b)}$ RA(2000) and DEC(2000), from {\sl RC3} (de Vaucouleurs et al. 1991).

$^{(c)}$ Distance, from Dahlem et al. (1997). 

$^{(d)}$ Total $B$-magnitude, corrected for Galactic and internal extinction, from {\sl 
RC3} (de Vaucouleurs et al. 1991).

$^{(e)}$ $B$-band luminosity, derived using the quoted distance and $B$-magnitude.

$^{(f)}$ Galactic HI column density (Dickey \& Lockman 1990).

$^{(g)}$ Far-infrared (FIR) luminosity (from Dahlem et al. 1997). The FIR flux is 
defined (Helou et al. 1985) as a combination of the {\it IRAS} $60\mu m$ and $100\mu 
m$ fluxes according to $f_{\rm FIR} \equiv 1.26 \times 10^{-11} (2.58\, f_{60} + 
f_{100})$ erg s$^{-1}$ cm$^{-1}$, where $f_{60}$ and $f_{100}$ are expressed in Jy.

$^{(h)}$ Inclination angle with respect to line of sight (Dahlem et al. 1998).

$^{(i)}$ 25-$B$-mag photometric radius, corrected for inclination and Galactic absorption,
derived using the angular value given by {\sl RC3} (de Vaucouleurs et al. 1991) and the 
quoted distance.

$^{(j)}$ 0.5-2 keV luminosity, derived using the quoted distance and the data 
presented in this paper. Not corrected for Galactic absorption.

$^{(k)}$ 2-10 keV luminosity, derived using the quoted distance and the data 
presented in this paper. Not corrected for Galactic absorption.

\end{table*}

In this paper we present the first broad-band (0.1-50 keV) spectral observations of 
NGC~4666, obtained using the {\it BeppoSAX} and {\it XMM-Newton} orbiting telescopes. In Sections 2 and 3 
we describe the data reduction and analysis; in Section 4 we describe the spectral 
fitting, whose implications are discussed in Sections 5 and 6. Our main results are 
summarized in Section 7. Unless explicitly quoted, all the errors reported in this paper 
are at the 90\% confidence level for one interesting parameter ($\Delta \chi^2 = 2.71$).

\section{Observations and Data reduction} 

\subsection{BeppoSAX measurements}

NGC~4666 was observed by {\it BeppoSAX} with the three narrow-field instruments (NFIs) (see 
Tables 2 and 3): the Low Energy Concentrator Spectrometer (LECS: Parmar et al. 1997), 
the Medium Energy Concentrator Spectrometer (MECS: Boella et al. 1997), and the 
Phoswich Detector System (PDS: Frontera et al. 1997).

\begin{table*}
\caption[]{ {\it BeppoSAX} and {\it XMM-Newton} instrumental parameters.}
\begin{flushleft}
\begin{tabular}{llllllllllllllll}
\hline
\hline
 & \multicolumn{3}{l}{FOV $^a$}   &\  &
        \multicolumn{3}{l}{$E_1-E_2$ $^b$}   &\  &
             \multicolumn{3}{l}{$\Delta E$ $^c$}   &\  &
                  \multicolumn{3}{l}{Sensitivity $^d$}   \\
\hline
\hline
{\sl SAX} & LECS & MECS & PDS &  & LECS & MECS & PDS &  & LECS & MECS & PDS &  & LECS & MECS & PDS \\
\cline{2-4}\cline{6-8}\cline{10-12}\cline{14-16}
 & $37^\prime$ & $56^\prime$ & $\sim$80$^\prime$ &   & 0.1--4.5 & 1.6--10 & 13--300 &   & 0.24 & 0.24 & 9 &  & 
                                   10$^{-13}$ & 10$^{-13}$ & 2$\times$10$^{-11}$ \\
\hline
\hline
{\sl XMM}  & MOS1/2 & PN & & & MOS1/2 & PN & & & MOS1/2 & PN & & & MOS1/2 & PN & \\
\cline{2-3}\cline{6-7}\cline{10-11}\cline{14-15} 
 & 30$^{\prime}$ & 30$^{\prime}$ & & & 0.3--10 & 0.5--10 & & & 0.13 & 0.13 & & & 7$\times$10$^{-15}$ & 
                   7$\times$10$^{-15}$ &\\ 
\hline
\hline
\end{tabular}
\end{flushleft}

$^a$ Diameter of the field-of-view.

$^b$ Instrument energy range, in keV.

$^c$ Energy resolution, in keV (at 6 keV for LECS, MECS, MOS1/2, PN; at 60 keV for PDS).

$^d$ Flux from point source after 100 ks exposure (erg cm$^{-2}$ s$^{-1}$).
\end{table*}

\begin{table*}
\caption[]{ {\it BeppoSAX} and {\it XMM-Newton} observation log.}
\begin{flushleft}
\begin{tabular}{lllllllllll}
\hline
\hline
  &  Date & & \multicolumn{3}{l}{Exposure time $^a$ (ks)} &\  &
                                    \multicolumn{3}{l}{Count Rate $^b$ ($10^{-3}$ ct s$^{-1}$)} \\
\hline
\hline
{\sl SAX} & 2001 Dec  14-18 &  & LECS & MECS & PDS &  & LECS & MECS & PDS &  \\
\cline{4-6}\cline{8-10}
 & & & 42.92 & 131.83 & 58.11 & & $2.23$ & $3.76$ & $<18.0$  \\
\hline
\hline
{\sl XMM} & 2002 June 27-28 &  & MOS1/2 & PN &     &  & MOS1/2 & PN & \\
\cline{4-5}\cline{8-9}
  &  &  & 57.1 & 48.3 & & & $\sim 0.013$ & $0.04$ & \\
\hline
\hline
\end{tabular}
\end{flushleft}

$^a$ On-source net exposure time. The LECS exposure time is considerably shorter than 
that of the MECS, because the LECS can operate only when the spacecraft is not 
illuminated by the Sun.

$^b$ Background-subtracted source count rates, with photon counting statistics errors. 
For the PDS the $90\%$ upper limit is given.

\end{table*}

The cleaned and linearized data produced by the {\it BeppoSAX} Scientific Data Center
	\footnote{\small See http://www.asdc.asi.it/bepposax} 
(ASDC) were analyzed using standard software packages (XSELECT v1.4, 
FTOOLS v4.2, and XSPEC v11.2). For the MECS, we used the event file made by 
merging the data of the two, properly equalized, MECS units. At the spatial 
resolution of the NFI instruments, NGC~4666 was not resolved. No significant 
source flux variation was detected over the observing period. 

\subsection{XMM-Newton measurements}

NGC~4666 was observed by {\it XMM-Newton} on 2002 June 27-28, for a total good exposure of 
$\sim$50 ks with the European Photon Imaging Camera (EPIC) detectors. Data were 
taken with the EPIC PN detector (Str\"uder et al. 2001) in extended full-frame 
mode, and with the EPIC metal oxide semiconductor (MOS) detectors (Turner et al. 
2001) in full-window mode (see Tables 2 and 3). The data were reduced using 
version 5.4.1 of the XMM-SAS software, using the standard processing scripts 
({\it emproc} and {\it epproc}). 10-12 keV light curves were inspected and 
found to be at a nominal rate for the whole observation. Data were selected 
using event patterns 0-12 (for the MOS) and pattern 0-4 (for the PN), and only 
good X-ray events (using the selection expression ``FLAG=0'' in {\it evselect}) 
were included. 

Images obtained with the PN detector and with the MOS1 are shown in Fig.1. 
These clearly show that NGC~4666 is extended in both energy bands. The 
extension is most prominent at the lower energy band, and seem to originate in 
a truly diffuse emission. At higher energies, the emission is primarily from 
point-sources, including a nuclear source that is the brightest.

\section{Spectral analysis}

\subsection{BeppoSAX data}

To maximize the statistics and the $s/n$ ratio, the LECS and MECS source counts 
were extracted from a circular region of $4^\prime$ radius. Background counts 
were extracted from high Galactic latitude ``blank'' fields (provided by the {\it BeppoSAX} 
ASDC) using an extraction region corresponding in size and detector position to 
that used for the source. The PDS spectra extracted with the standard pipeline 
(with the rise-time correction applied) provided directly by the {\it BeppoSAX} ASDC; the 
simultaneously measured off-source background was used. Significant emission of 
NGC~4666 was detected at significant levels by the LECS and MECS, but not by 
the PDS instrument (see Table 3).

Spectral channels corresponding to energies 0.1-4 keV and 1.8-10 keV have been 
used respectively for the analysis of the LECS and MECS data [as suggested by 
the {\it BeppoSAX} `Cookbook' (see Fiore et al. 1999)]; PDS data are in the 13-60 keV 
spectral channels. LECS and MECS source counts were rebinned to have $s/n>3$ in 
each energy bin; PDS source counts were rebinned to have $s/n>2$. Standard 
calibration files were used in the fitting procedure; the LECS/MECS and PDS/MECS 
cross-constants normalizations were allowed to vary in the ranges proposed by the 
{\it BeppoSAX} `Cookbook'. 

\subsection{XMM-Newton data}

The nuclear spectrum was extracted (for both MOS and PN detectors) from a circular 
region of 15$^{\prime\prime}$ radius around the nucleus (shown in Fig.1).
Background spectra were extracted from offset circles, close to NGC~4666 but free 
of any background source. Standard response matrices and ancillary files were used. 
The background-subtracted spectrum was fitted using XSPEC version v11.2, including 
0.3-10 keV data for the MOS and 0.5-10 keV data for the PN. Data were binned so as 
to have at least 20 counts/bin to enable the use of $\chi^{2}$ minimization during 
the spectral fits. The plot in Fig.2 was obtained using $s/n \leq 3$ (per bin). 

\begin{figure}
\vspace{10.95cm}
\includegraphics{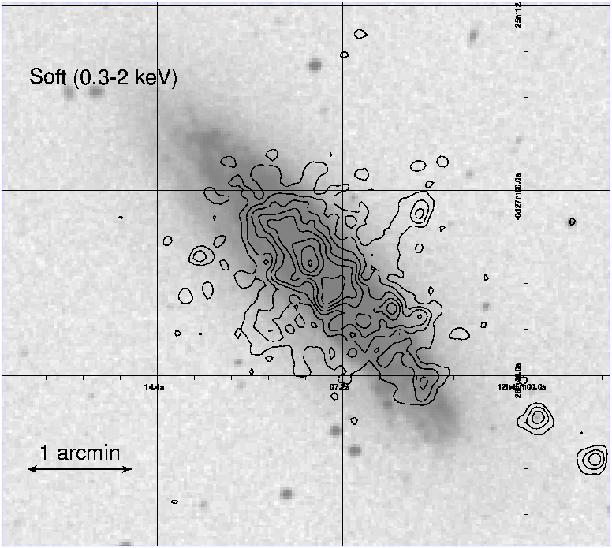}
\vskip 0.5cm 
\includegraphics{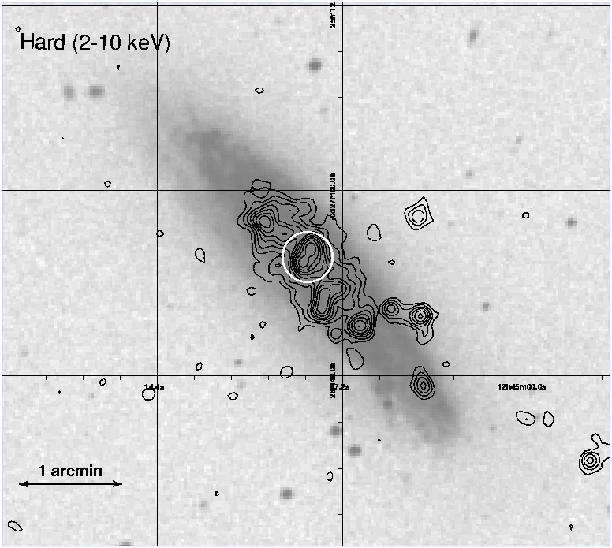}
\caption{Soft ({\it top}) and hard ({\it bottom}) X-ray contours obtained with the 
MOS1 overplotted on the optical image from the DSS. Contours have been obtained 
after gaussian smoothing. They span a surface brightness range of $\sim$ 0.38--9.4 
c arcsec$^{-2}$ in the soft band, and $\sim$ 0.3--3.6 c arcsec$^{-2}$ in the hard 
band. In the lower panel, the 15$^{\prime\prime}$-radius annulus marks the nuclear 
extraction region, data from which are shown in Fig.5.
}
\end{figure}

\section{Main spectral components}

The basic model components used in various combinations to fit {\it BeppoSAX} data are a PL, a 
gaussian line, and the Mekal model describing thermal emission from optically thin hot 
plasma (see the XSPEC package). We include photoelectric absorption due to Galactic 
foreground, corresponding to HI column density $N_{\rm H,Gal}=1.7 \times 10^{20}$ 
cm$^{-2}$. Also included is instrinsic absorption in NGC~4666 to be determined from the 
spectral analysis. The results of the spectral analysis are summarized in Table 4.

\begin{table}
\caption[]{ Spectral analysis of {\sl BeppoSAX} data}
\begin{flushleft}
\begin{tabular}{  l    l }
\hline
 Model          &  Parameters\\
\hline

{\bf A}: pl     &  $\chi^2/\nu$ = 49.62/43   \\
$\Gamma $    &  $2.15 \pm 0.15$  \\
\hline

{\bf B}: mekal & $\chi^2/\nu$ = 50.24/43  \\
$kT$ (keV)        & $4.09 \pm 0.77$   \\
$Z$ $(Z_\odot)$   & $1.0$ (f)  \\
\hline

{\bf C}: pl+gauss & $\chi^2/\nu$ = 38.63/41 \\
$\Gamma $    & $2.27 \pm 0.17$ \\
Line: $\s_{\rm E}$; E(keV); EW (keV)   &  $0.0$ (f); $6.46 \pm 0.55$;  $1.87 \pm 0.60$ \\
\hline

{\bf D}: 
mekal+pl+gauss & $\chi^2/\nu$ = 33.46/39 \\
$kT$ (keV)        & $0.23 \pm 0.05$   \\
$Z$ $(Z_\odot)$   & $1.0$ (f)  \\
$\Gamma $    & $2.13 \pm 0.19$ \\
Line: $\s_{\rm E}$; E(keV); EW (keV)   &  $0.0$ (f); $6.48 \pm 0.11$;  $1.66 \pm 0.50$ \\
\hline

\end{tabular}
\end{flushleft}
\end{table}

Single-component models do not provide an adequate description of the {\it BeppoSAX} 
broad-band spectrum of NGC~4666. A single unabsorbed PL model (model A) and a 
single-temperature thermal model (model B) are rejected at $>$99.9$\%$ confidence 
level. In Fig.2-{\it left} we show the ratio between the data and model A. A 
broad excess is evident at $\sim$0.5 keV, and a line-like excess is seen at 
$\sim$6.5 keV. The residuals around 0.5 keV suggest the presence of a soft 
thermal component, and the feature at $\sim$6.5 keV suggests Fe-K$\a$ line 
emission. 

Multi-component models including different combinations of thermal, PL, and 
gaussian-line components can then be fitted to the data. A PL+gaussian-line 
model (model C), although clearly marking an improvement over models A and B, is 
still unsatisfactory because it does not remove the excess residuals at $\sim$0.5 keV 
(Fig.2-{\it right}). A simple model that provides an acceptable statistical fit 
(model D) includes: {\it i)} an unabsorbed thermal plasma with $kT \simeq 0.23 
\pm 0.05$ keV (with the chemical abundance set to the solar value); {\it ii)} 
an unabsorbed PL with photon index $\G \simeq 2.1 \pm 0.2$; and {\it iii)} a 
narrow ($\s_{\rm E} \sim 0$) gaussian line at $E = 6.47 \pm 0.12$ keV, with 
equivalent width EW $\sim$ 1.7 keV. In Fig.3 we show model D ({\it top left}, with 
unfolded instrumental response) and the residuals of the fit ({\it bottom left}), 
as well as the confidence countours of some physically meaningful parameter 
pairs ({\it right}). [If we allow for intrinsic absorption, none is deduced in the soft thermal 
component, while some is found (at $>90\%$ confidence level) in the PL component, 
corresponding to $N_{\rm H} = (1.0 \pm 0.8) \times 10^{22}$ cm$^{-2}$. Intrinsic 
absorption would cause the PL slope to steepen to $\G = 2.7 \pm 0.5$; see Fig.3-{\it middle
right}.]

The spatially integrated 0.5-10 keV spectra of local SBGs are well fitted by a 
model including a soft ($kT < 1$ keV) thermal emission, and a higher energy 
component consisting of either a $\magcir$5 keV thermal emission, or a $\G \sim 
1.5 - 2$ PL (e.g., Dahlem et al. 1998). By its continuum properties, therefore, 
the {\it BeppoSAX} spectrum of NGC~4666 seems to be a typical SBG spectrum.

\begin{figure}
\vspace{3.0cm}
\includegraphics{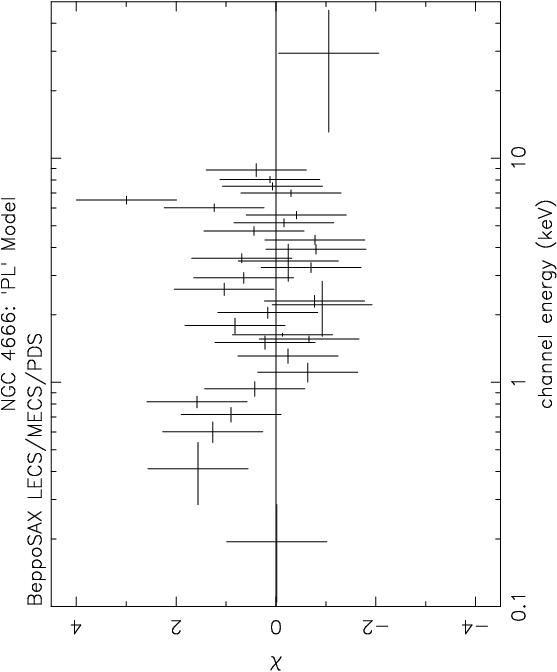}
\hskip 0.3cm
\includegraphics{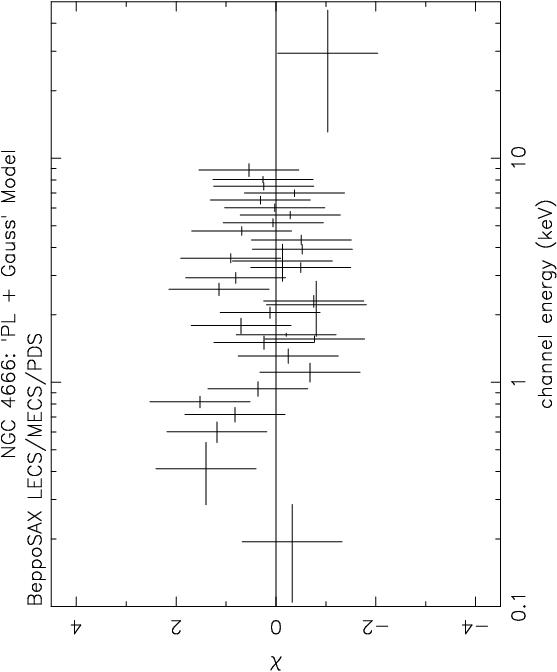}
\caption{
{\it Left}: residuals of the fit of BeppoSAX data to model A (PL with $\G=2.15$). 
{\it Right}: residuals of the fit of BeppoSAX data to model C (PL with $\G=2.27$,
plus a gaussian line at $E =6.46$ keV): the excess at $\sim$6.5 keV has disappeared, 
but not that at $\mincir$1 keV.
}
\label{fig:singlefit}
\end{figure}

\begin{figure}
\vspace{8.0cm}

\includegraphics{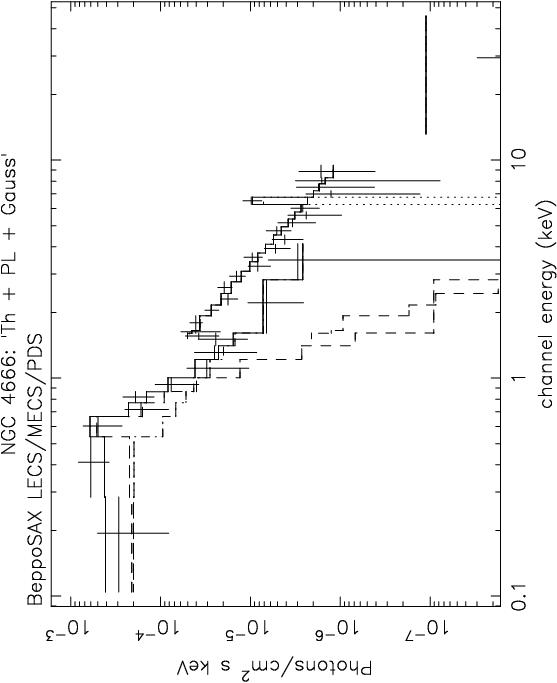} 
\includegraphics{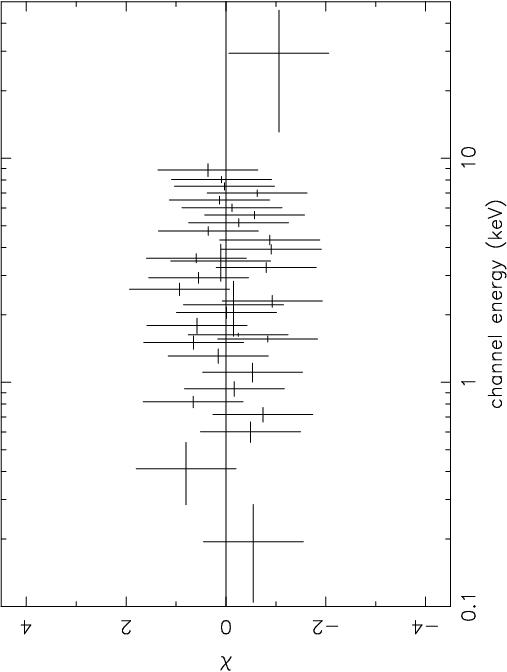}
\includegraphics{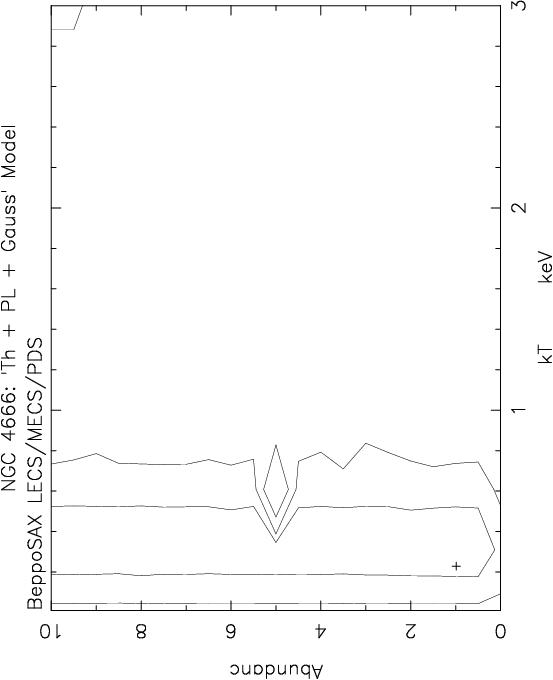}
\includegraphics{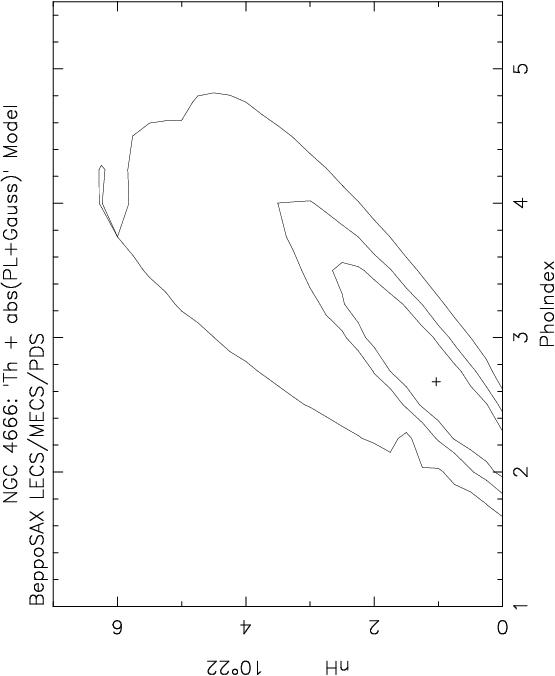}
\includegraphics{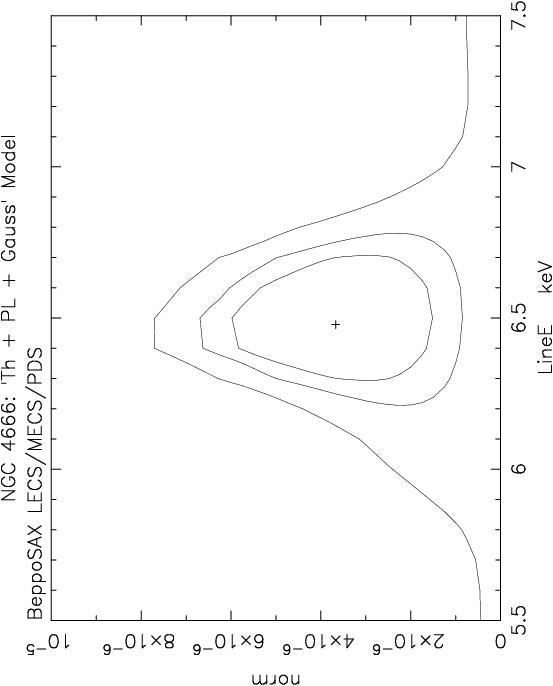} 
\vspace{0.2cm}
\caption{
The model spectrum C, unfolded of instrumental response ({\it top left}); and 
the fit residuals ({\it bottom left}).
The 68\%, 90\% and 99\% probability contours for two interesting parameters for:
{\it (top right)} the temperature, $kT$ (in keV), and the chemical abundance, 
	$Z$ (in solar units), of the thermal component; 
{\it (middle right)} the intrinsic photon index, $\G$, and the {\it in situ} 
HI column density, $N_{\rm H}$, for the PL component; and {\it (bottom right)} 
for Fe-K line energy vs. normalization. 
}
\end{figure}

\section{The Fe-K$\a$ emission line}

What makes the integrated spectrum of NGC~4666 quite outstanding among known SBG 
spectra is the huge (EW $\sim$ 2 keV) emission feature clearly required by the 
{\it BeppoSAX} data at $\sim$6.5 keV (see Fig.3, bottom right). In fact, in SBG spectra 
emission between 6 and 7 keV is uncommon, inconspicuous (EW $\mincir 0.3$ keV), 
and centered at 6.7 keV (and hence it is identified with K$\a$ emission from 
high-ionization iron -- e.g., Persic et al. 1998; Griffiths et al. 2000; Pietsch 
et al. 2001). 

As this emission is unresolved in the {\it BeppoSAX} data, we search 
for it in the archival {\it XMM}-Newton data. The superior spatial resolution of 
the latter allows us to identify the line as originating from the {\it nuclear} 
region, and clearly distinguish it from the {\it extended} continuum emission 
(see Fig.4). 

The nuclear location of the line, revealed by {\it XMM-Newton}, could be the clue to the presence 
of a strongly absorbed AGN in NGC~4666. If so, according to the Compton-thick model 
predictions we expect to observe a prominent (EW $\magcir 1$ keV) line centered at 
6.4 keV and superposed on a flat continuum underlying the line itself and cospatial 
with it. 

Spectral analysis of {\it XMM}-Newton data for the nuclear and circum-nuclear 
region does indeed show such a line ($E \simeq 6.42 \pm 0.02$ 
keV with EW $\simeq 0.7 \pm 0.2$ keV), superimposed on a flat ($\G \sim 1.3$) 
continuum (see Fig.5 and Table 5). We identify the line as due to the 
K$\a$ transition at 6.4 keV from low-ionization iron. 

To check for consistency of the Compton-thick hypothesis, we try a cold 
continuum reflection model (PEXRAV in the standard XSPEC spectral fitting 
package).  This model represents the emission that would be expected 
from, e.g., a circum-nuclear torus that totally blocks the direct view 
of the primary AGN spectrum (i.e., the type-2 configuration in the 
unified model of AGNs). As commonly done for standard primary spectra 
of Sy1 and Sy2 galaxies (e.g. Turner et al. 2000), we assume 
a PL with photon index $\G=1.9$ for both the primary (completely absorbed) 
component and the warm-scattered component (which is visible in the $\sim$ 
1-5 keV band). We also assume solar chemical abundances, full ionization of 
H and He, and a viewing angle of 80$^\circ$ (i.e., we take the torus axis to 
coincide with the spin axis). The resulting fit of this (astrophysically 
motivated) model is satisfactory (see Table 5), similar to the 
previous case of the (heuristic) single-PL model. The 2-10 keV 
flux of the best-fit cold-reflection component is $F_{\rm refl}^{\rm AGN} 
\sim 4.8 \times 10^{-14}$ erg cm$^{-2}$s$^{-1}$, yielding an observed luminosity 
of $L_{\rm refl}^{\rm AGN} \sim 4.0 \times 10^{39}$ erg s$^{-1}$ (see Fig.5). 
If we assume -- as did Iwasawa et al. (1997) -- that the intrinsic luminosity 
of the hidden AGN is a factor $\magcir$50 larger than the observed value, then 
$L_{\rm intr}^{\rm AGN} \magcir 2 \times 10^{41}$ erg s$^{-1}$.    

It should be emphasized that the extraction radius ($\sim$15$^{\prime\prime}$) 
corresponds to a linear size of 1.9 kpc. Thus, this large circum-nuclear region includes a 
substantial fraction of the SB emission. This may well explain why the continuum underlying 
the 6.4 keV line is somewhat steeper than predicted by a pure Compton-thick model (see Fig.2). 

\begin{table}
\caption[]{ Spectral analysis of {\sl XMM-Newton} data}
\begin{flushleft}
\begin{tabular}{  l    l }
\hline
 Model          &  Parameters\\
\hline

{\bf A}: mekal+pl+gauss     &  $\chi^2/\nu$ = 220.05/163   \\
$kT$ (keV)        & $0.33 \pm 0.02$   \\
$Z$ $(Z_\odot)$   & $1.0$ (f)  \\
$\Gamma $    & $1.31 \pm 0.05$ \\
Line: $\s_{\rm E}$; E(keV); EW (keV)   &  $0.0$ (f); $6.42 \pm 0.05$;  $0.75 \pm 0.25$ \\
\hline

{\bf B}: mekal+pl+gauss+refl. & $\chi^2/\nu$ = 236.35/163  \\
$kT$ (keV)         & $0.32 \pm 0.02$   \\
$Z$ $(Z_\odot)$    & $1.0$ (f)  \\
$\Gamma $          & $1.9$ (f) \\
$R$ (= refl./dir.) & $6 \pm 2$ \\
Line: $\s_{\rm E}$; E(keV); EW (keV)   &  $0.0$ (f); $6.42 \pm 0.05$;  $0.75 \pm 0.25$ \\
\hline

\end{tabular}
\end{flushleft}
\end{table}

\begin{figure}
\vspace{7.5cm}
\includegraphics{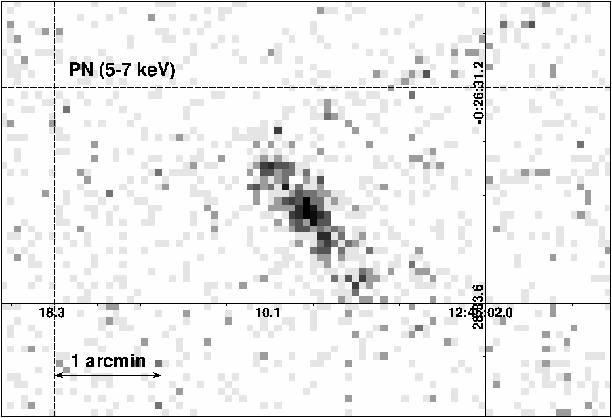}
\vspace{0.5cm}
\includegraphics{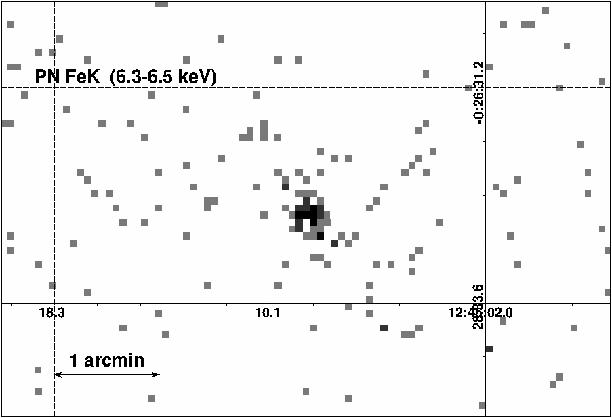}
\vspace{0.19cm}
\caption{PN raw image (1pix = 4 arcsec) in the 5-7 keV energy band ({\it 
top}), and in the 6.3-6.5 keV band ({\it bottom}). 
}
\end{figure}

\begin{figure}
\vspace{4.0cm}
\includegraphics{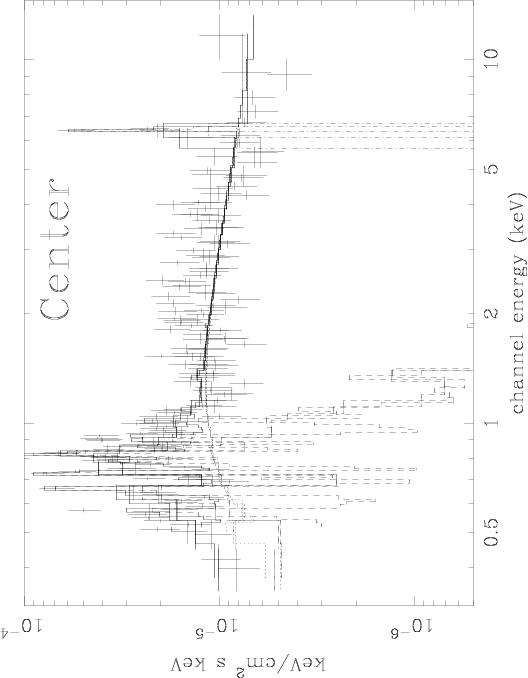}
\hskip 0.25cm
\includegraphics{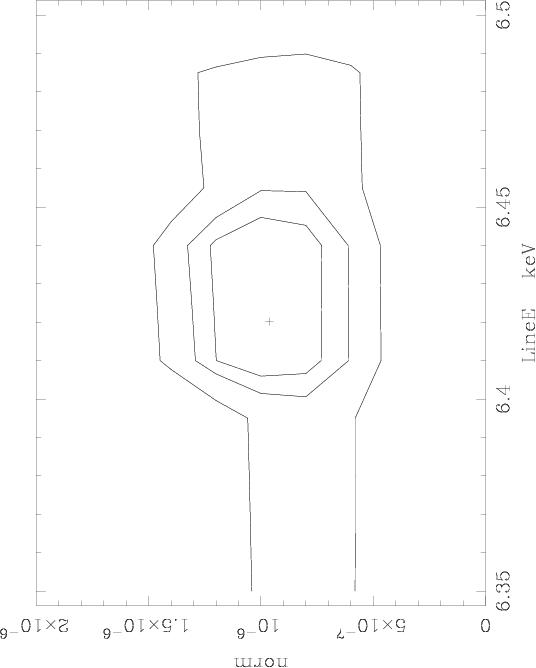}
\caption{
{\it Left}: Spectrum from the nuclear (15$^{\prime\prime}$ radius) region (see Fig.1, {\it
            bottom}). A very strong neutral Fe-K$\a$ line at 6.40 keV, and the underlying 
	    cospatial flat continuum (approximated by a $\G \simeq 1.3$ PL model), are 
	    clearly visible. 
{\it Right}: 68\%, 90\% and 99\% probability contours for the Fe-K$\a$ line's energy versus 
	    normalization. The emission line is clearly significant and spectrally localized.
}
\end{figure}

\section{Origin of the continuum}

\subsection{Thermal component}

The thermal plasma component identified in the spatially integrated {\it BeppoSAX} data 
agrees (within errors) with the earlier finding (based on \rosat data) by Dahlem 
et al. (1998). Our {\it BeppoSAX} data are not sensitive enough to determine the chemical 
abundance, whereas the temperature is better constrained to $kT \sim 0.3 \pm 0.1$ 
keV for a broad range of the abundance (see Fig.3, upper right); also, the temperature 
is decoupled from the slope of the PL component. 

Thermal emission is usually interpreted as arising from the interaction between 
the outgoing galactic wind and the ambient ISM. The wind is hot but very tenuous, 
while the ISM is dense but cold, so none of these is likely to emit appreciably 
in the X-ray region. The boundary region between the two phases, however, does 
achieve a combination of temperature and density suitable for sub-keV emission 
from a few percent of the wind mass (Strickland \& Stevens 2000; Suchkov et al. 
1996). Single-temperature models ($kT \sim 0.8$ keV) and sometimes two-temperature models 
(with $kT \sim 0.8$ and $\sim 0.3$ keV) have been deduced from \asca + \rosat 
(e.g., Dahlem et al. 1998) and {\it XMM-Newton} (Franceschini et al. 2003) data of actively 
star-forming galaxies. 
	\footnote{ Part of the measured thermal emission may also come from 
	OB stars, which are expected to be abundant in SB regions and 
	whose thermal emission is characterized by $kT \sim 0.2$ keV, 
	$L \sim 10^{33}$ erg s$^{-1}$ (see PR02). }

Analysis of archival {\it XMM-Newton} data shows that the soft (0.3-2 keV) emission is diffuse 
and essentially co-extended with the higher energy 2-10 keV emission (see Fig.1). 
Because the soft emission is largely dominated by the thermal component, which arises 
from the galactic winds that are driven by current SF activity, its spatial distribution 
suggests that the SB occurs throughout the disk in NGC~4666.

\subsection{PL component}

Persic \& Rephaeli (2002; hereafter, PR02) have quantitatively assessed the 
roles of the various X-ray emission mechanisms in star-forming galaxies. They 
have used an equilibrium stellar-population synthesis model of the Galactic 
population of high-mass X-ray binaries (HMXBs) and low-mass X-ray binaries 
(LMXBs). The abundance of SN remnants (SNRs), from both Type II 
and Type Ia events, was also consistently estimated. From the literature PR02 
derived typical spectra for these source classes. The spectral properties and 
relative abundances of the various classes of stellar sources allowed PR02 to 
calculate the composite X-ray spectrum arising from a stellar population of 
Galactic composition.

PR02 determined that the $\sim$2-15 keV emission is dominated by LMXBs; this is 
not surprising, given that the synthetic spectrum was calibrated by that of a 
normal, quietly star forming galaxy like ours. Due to the long times between the 
formation and the onset of the X-ray phase (the optical companion is a subsolar 
main-sequence star) in LMXBs, their emission traces the higher, {\it past}
Galactic SFR, and not the lower, {\it current} SFR. Thus in order to obtain an 
estimate of the current SFR, the spectral contribution of LMXBs should not be 
included. 

Recent observations of nearby galaxies -- mostly with {\it Chandra} -- have 
revealed two new populations of X-ray point sources (XPs), not present in our 
Galaxy, with 2-10 keV luminosities in the range $2 \times 10^{38}$ erg s$^{-1}
\leq L_{2-10} \leq 5 \times 10^{39}$ erg s$^{-1}$ (Very Luminous Sources: 
VLXs), and with $L \geq 10^{39}$ erg s$^{-1}$ (Ultra-Luminous Sources: ULXs) 
(Fabbiano \& White 2003)
	\footnote{ The two quoted limits correspond to 
        Eddington luminosities for, respectively, a $1.5
	\,M_\odot$ neutron star and for a $8\, M_\odot$ 
	black hole (BH) which is the limiting 
	BH mass obtainable via ordinary stellar evolution. 
	ULXs are called 'super-Eddington sources'.}.
In the 2-10 keV band relevant here, mean VLX and ULX spectra are best-fit 
by PLs with photon indexes of $\sim$2 and $\sim$1.2, 
reminiscent of, respectively, Galactic Black-Hole-X-ray binaries (BHXBs) 
	\footnote{In a few cases the optical counterparts 
	of ULX have been identified as O stars (Liu et al. 
	2002; Roberts et al. 2002b), thus suggesting a 
	relation between ULXs and HMXBs, which of course 
	are young objects that trace the current SFR.}
(e.g., Zezas et al. 2002; Terashima \& Wilson 2003; Foschini et al. 2002; see also 
Fabbiano \& White 2003). VLXs and ULXs may describe different luminosity/spectral
states of BHXBs as a function of the mass infall rate onto the BH
	\footnote{ Qualitatively, the inverse correlation between 
	$L_{2-10}$ and $\G$ (i.e., flatter spectra for higher luminosities)
	can be understood within the standard accretion model devised to 
	explain the high $L_{\rm x}$ of interacting  binaries. The increase
	in $L_{\rm x}$ is driven by an increase of $\dot M$, which in turn 
	means a piling up of material around the emission region that leads 
	to a higher Compton scattering optical depth, and hence to a flatter 
	spectrum. E.g., according to Sunyaev \& Titarchuk (1980) and Shapiro 
	et al. (1976), if a source of photons is embedded in a plasma cloud 
	of optical depth $\tau$ and temperature $T$, the escaping radiation 
	has a high-energy photon index 
$$
	\Gamma ~=~ \biggl[ {9 \over 4} ~+~ {\pi^2 m_{\rm e} c^2 \over 
	3 \, (\tau + 2/3)^2 \, kT} \biggr]^{1/2} ~-~ {1 \over 2} \,.
$$
	}. 
The discovery of these sources has augmented the variety of known stellar endproducts 
that emit X-rays, and the PR02 method can easily incorporate such new classes (Persic 
\& Rephaeli 2004). 

The population of VLX and ULX sources in a galaxy is more abundant the more active is 
star formation in that galaxy. Specifically, if the integral XP luminosity function 
(XPLF) is described as $N(>L) \propto L^{-\alpha}$ (which corresponds to a differential 
XPLF of the type ${dN \over dL} \propto L^{-(1+\alpha)}$), based on a wealth of {\it 
Chandra} oservations it has been found that XPLFs are less steep in SFGs than in normal 
spirals and ellipticals. (For example, the two most nearby SB galaxies, have $\alpha 
\sim 0.5$ (M~82) and $\alpha \sim 0.8$ (NGC~253), while normal spirals have $\alpha \sim 
1.2$, and ellipticals have approximately $\alpha \geq 1.4$; see Kilgard et al. 2002.) So
the integrated point source luminosity of SFGs is dominated by the highest-$L$ sources 
(Colbert et al. 2003). If the XPLF in NGC~4666 is similar to those in other star-forming 
galaxies (e.g., $\a \sim 0.5-0.8$; see Kilgard et al. 2002), then its spatially integrated 
2-10 keV luminosity and spectrum will be dominated by VLXs and ULXs (unless there is a 
substantial diffuse hard component). The resulting spectral index, $\bar\G$, will be $1.2 
\mincir \bar\G \mincir 2$, with the exact value depending on the XPLF index $\a$ and on 
$\G = \G(L)$. Data with angular resolution at the arcsec level and sensitivity at the 
$\sim$10$^{-15}$ erg cm$^{-2}$s$^{-1}$ level (e.g., \chandra data) will resolve XPs in 
NGC~4666 down to $L_{2-10} \sim 5 \times 10^{37}$ erg cm$^{-2}$ s$^{-1}$. The fluxes and 
-- where possible -- the spectra obtained for these XPs would allow one to obtain some 
firmer clues on their nature. A full analysis of the maxima of emission seen in the 2-10 
keV {\it XMM-Newton} map is beyond the scope of this paper. A preliminary analysis is, however, in order. 

\begin{figure}
\vspace{6.0cm}
\includegraphics{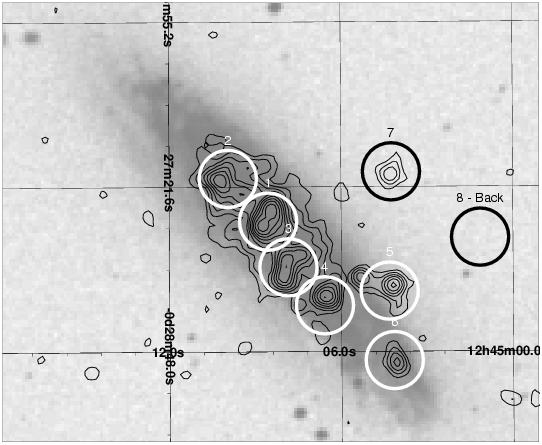}
\caption{2-10 keV contours obtained with the MOS1 overplotted on the optical image from 
the DSS (for details see Fig.1). The seven "sources" (i.e., maxima of emission) are 
identified by $15^{\prime \prime}$-radius circles. An eighth circle denotes an empty 
field that was used for background estimation.
}
\end{figure}

From the 2-10 keV isophotes overlaid on the optical image of NGC~4666 (see Fig.6), 7 emission 
maxima can be identified by visual inspection. In order to estimate the fluxes from these 
locations, we have constructed 7 circles, with radii of $15^{\prime \prime}$, centered on the 
selected features (as to include $\geq$70$\%$ of the encircled energy). An eighth, off-galaxy 
circle was used to estimate the background. Assuming for each "source" a flat spectral shape 
($\G=1.3$) in order to obtain a (likely) upper limit to its 2-10 keV flux and a background 
as given by the off-galaxy empty field, the most obvious conclusion was that 
"sources" \#1-4 qualified as ULXs and "sources" \#5-7 qualified as VLXs. 
However, of such sources in Fig.6, 
three quite clearly show structure: \#1, \#3 and \#6.
These emissions can be interpreted as resulting from (at least) 2, 
3, and 2 sources each, respectively. In fact, concerning emission \#1, 
we know that about half of the emission is due to "cold reflection" from a deeply absorbed 
type-2 AGN ($L_{2-10}^{\rm AGN, refl} \sim 0.40 \times 10^{40}$ erg s$^{-1}$, see section 
5): the remaining flux, corresponding to $L_{2-10} \sim 3.8 \times 10^{39}$ erg cm$^{-2}$ 
s$^{-1}$, can be attributed to an ULX. As for emission \#3, 
that may be composed 
of 3 VLXs with $L_{2-10} \sim 10^{39}$ erg cm$^{-2}$ 
s$^{-1}$ each. Finally, the elongated emission \#6 
may come from 2 VLXs with $L_{2-10} \sim 5 
\times 10^{38}$ erg cm$^{-2}$ s$^{-1}$ each. Furthermore, emission \#2
looks quite confused and its isophotes, which tend to become more triangular 
for deacreasing levels of emission, may suggest the presence of three unresolved sources: 
given the global level of emission measured for this feature, such three sources could be 
(e.g.) 2 VLXs with $L_{2-10} \mincir 10^{39}$ erg s$^{-1}$ each, and 1 ULX with $L_{2-10} 
\magcir 10^{39}$ erg s$^{-1}$. 

Based on these considerations, we {\it tentatively} identify the XPs hinted at by the 2-10 
keV {\it XMM-Newton} image as being 3 ULXs (with luminosities $2 \mincir L_{2-10}/(10^{39}$ erg s$^{-1}) 
\mincir 4$) and 9 VLXs (with luminosities $0.5 \mincir L_{2-10}/(10^{39}$ erg s$^{-1}) 
\mincir 2$). Collectively, they contribute $1.6 \times 10^{40}$ erg cm$^{-2}$ s$^{-1}$ 
in the 2-10 keV band. Adding the estimated cold reflection from the hidden AGN, the total 
XP luminosity of NGC~4666 becomes $L_{2-10}^{\rm XP} = 2 \times 10^{40}$ erg s$^{-1}$, 
i.e. $80\%$ of the total 2-10 keV luminosity observed with {\it BeppoSAX} (the latter being $L_{2-10} 
= 2.5 \times 10^{40}$ erg cm$^{-2}$ s$^{-1}$, see Table 1). The remaining $20\%$, missed by 
the XP luminosity computed from the {\it XMM-Newton} map as described above, might be genuinely diffuse 
(see below), or it could be due to unresolved sources of lower luminosity (i.e., with $L_
{2-10} \mincir 5 \times 10^{38}$ erg s$^{-1}$), or a combination of the two. 

We can gain an insight into this issue by the following argument. If the differential XPLF is 
of the type ${{\rm d}N \over {\rm d}L} \propto L^{-(1+\alpha)}$, our tentative identification 
of XPs suggests $\a \sim 0.5$ (similar to what is observed in M~82, see Kilgard et al. 2002): such 
XPLF implies that the collective observed luminosity due to our tentative XPs amounts to $76\%$ 
of the XP luminosity integrated down to $10^{38}$ erg s$^{-1}$ (i.e. down to the lower luminosity 
limit defining VLXs). This simple model implies that the XP luminosity, integrated between $10
^{38}$ erg s$^{-1}$ and $4 \times 10^{39}$ erg s$^{-1}$, is $L_{2-10}^{\rm XP} = (1/0.76 \times 
1.6 + 0.4 = 2.5) \times 10^{40}$ erg s$^{-1}$. This matches the integrated 2-10 keV luminosity 
seen with {\it BeppoSAX}. What about the spectral shape? If we assume VLXs to have $\G=2.0-2.5$ and ULXs 
to have $\G=1.3$, our toy model implies $\bar \G \simeq 1.7-2.0$, in reasonable agreement with
the {\it BeppoSAX} result $\bar \G = 2.1 \pm 0.2$. Therefore a toy model for the XP population, that 
involves ULXs and VLXs distributed according to an $\a = 0.5$ XPLF plus a Compton-thick AGN, 
can bring {\it BeppoSAX} data and {\it XMM-Newton} data into reasonable agreement as far as both the 
luminosity and spectral properties are concerned. 

That the 2-10 keV emission of NGC~4666, once the AGN emission has been 
subtracted, is indeed mostly related to current SF activity can be checked 
as follows. From a sample of SB-powered ultra-luminous IR galaxies (ULIRGs; 
Franceschini et al. 2003) it can be deduced that the 2-10 keV to FIR emission 
ratio of pure SBs is $\theta \equiv$ log$(L_{2-10}/L_{\rm FIR}) \sim -4.3 \pm 
0.3$ (Persic et al. 2004). Upon subtraction of the AGN-related emission, 
NGC~4666 has $\theta \sim -4.0$: so its disk 2-10 keV emission is roughly at 
the level implied by its current SFR. This deduction seems reasonable because 
NGC~4666 exhibits the "smoking gun" proof of being a global SB: a large-scale 
supergalactic wind and a substantial radio halo. It should also be recalled 
that NGC~4666 is member of a small interacting group of galaxies (Garc{\'\i}a 
1993); such a dense environment may well have tidally triggered a global SF 
activity in NGC~4666.

\subsubsection{Compton scattered radiation}

As is well known, Compton scattering of (the radio-synchrotron emitting) 
relativistic electrons by the FIR and cosmic microwave background 
(CMB) radiation fields results in a 
PL component with roughly the same photon index, $\G \sim 1.8$, as that 
of the radio emission (Schaaf et al. 1989; Rephaeli et al. 1991; Dahlem 
et al. 1997; Sukumar et al. 1988). This closely resembles the observed 
2-10 keV spectral profile. Compton emission may contribute $\mincir 20\%$ 
of the observed 2-10 keV flux, as suggested by the following argument. 
[Formally, this is the level of 2-10 keV emission needed to supplement
$L_{2-10}^{\rm XP}$ (as measured by {\it XMM-Newton}) in order to reproduce $L_{2-10}$ 
(as measured by {\it BeppoSAX}).] If the relativistic electrons and seed FIR photons 
are cospatial with a 
uniform magnetic field $B$, the Compton luminosity can be estimated from 
$L_{\rm c}  ~=~  L_{\rm s} ~ U_{\rm ph}/U_{\rm B}$, where $L_{\rm s}$ is 
the total synchrotron luminosity, $U_{\rm ph}$ is the energy density in 
the seed photon field, and $U_{\rm B}$ is the energy density in the 
magnetic field. Integrating the radio spectrum over the 0.01-200 GHz range 
(the upper limit of the integration contributes very little to the integral), 
we get $L_{\rm s} \simeq 5.1 \times 10^{39}$ erg s$^{-1}$. For a spherical 
region of radius $r$, the seed photon energy density is $U_{\rm ph} = 
3 L_{\rm bol} /(4 \pi r^2 c)$. Assuming $L_{\rm bol} \simeq L_{\rm FIR}$ and 
$r \simeq 3$ kpc (see Dahlem et al. 1997), we have $U_{\rm ph} \sim 2 \times 
10^{-11}$ erg cm$^{-3}$. If $B \simeq 15$ $\mu G$ in the disk where the FIR 
emission originates (Dahlem et al. et al. 1997), then $U_{\rm B} = B^2/(8 \pi) 
\sim 10^{-11}$ erg cm$^{-3}$. The Compton luminosity of NGC~4666, integrated 
over the entire range of upscattered energies is expected to be $L_{\rm c} 
\sim 10^{40}$ erg s$^{-1}$, suggesting that FIR-Compton emission may account 
for a non-negligible fraction of the observed 2-10 keV emission, $L_{2-10} = 
2.5 \times 10^{40}$ erg s$^{-1}$. However, this estimate depends sensitively 
on the mean value of $B$ which is not known very well. The corresponding 
estimate of emission from Compton scattering off the CMB (mostly) in the halo 
is even less secure, due to substantial uncertainty in the mean halo value of 
B (even ignoring pertinent spatial considerations), but is likely to be weaker 
than that of the disk.

\section{Conclusions} 

The X-ray spectrum of the nearly edge-on superwind galaxy NGC~4666 was measured 
in the 0.1-10 keV band with the LECS, MECS, and PDS instruments aboard {\it BeppoSAX}, as 
well as in the 0.3-10 keV band with the EPIC instrument aboard {\it XMM-Newton}. As seen by 
{\it BeppoSAX}, the emission is not spatially resolved, and significant detection occurs 
only at energies $<10$ keV. Emission measured by {\it XMM-Newton} is spatially resolved as 
coming from the whole disk. 

SB and AGN activities coexist in NGC~4666. The former, which extends over most of 
the disk, is revealed, in the X-ray region, by {\it (i)} diffuse thermal emission 
associated with SN-powered galactic wind, and by {\it (ii)} high-luminosity XPs 
that are well-known indicators of enhanced star-formation activity. It is also 
indicated by {\it (iii)} an XPLF index $\a \sim 0.5$, suggested here by a preliminary 
XP analysis of the {\it XMM-Newton} data and similar to those derived for other actively 
star-forming galaxies. The AGN activity is revealed by a prominent K$\a$ line 
from "cold" iron at 6.40 keV and its relatively flat underlying continuum, both 
coming from the nuclear region, that originate from the reflection of the primary 
continuum by the cold inner wall of the circumnuclear torus.

More specifically, our results can be summarized as follows.

\noindent
{\it (a)} At energies $\mincir 1$ keV the spectrum of NGC~4666 is dominated by 
a $\sim 0.25$ keV diffuse thermal plasma component (of unconstrained chemical 
abundance), distributed throughout the disk, which traces the interaction between 
the hot, tenuous wind elements and the cool, denser ISM in the disk. The 
distribution of this sub-keV plasma shows that the current SB phase involves the 
whole galaxy. 

\noindent
{\it (b)} At energies $\sim$2-10 keV the integrated spectrum is dominated by a
steep ($\G \sim 2$) PL component. This emission is probably due to unresolved
XPs distributed according to a luminosity function with (differential) index
$\a \sim 0.5$, and having $L_{2-10} \geq 10^{38}$ erg s$^{-1}$. These sources,
which are likely to be accreting binaries with $\leq$8$M_\odot$ black hole
components, are known to dominate the XP luminosity of nearby star-forming
galaxies and in the relevant energy range their spectra have $\G \sim 2$ PL
(for $0.1 \leq L_{2-10}/(10^{39}$ erg s$^{-1}) < 2$) and $\G \sim 1.3$ PL
(for $L_{2-10} \geq 2 \times 10^{39}$ erg s$^{-1}$). A $\G \sim 1.8$ PL
contribution from Compton scattering of relativistic electrons (whose
observed radio index is $\a_r \sim 0.8$) by the ambient FIR photons may
add a truly diffuse 2-10 keV emission.

\noindent
{\it (c)} The prominent (EW $\sim 2$ keV) emission line, seen but not resolved
in the {\it BeppoSAX} data, is identified in the {\it XMM-Newton} data as fluorescent neutral Fe-K$\a$
at 6.40 keV originating from the nucleus of NGC~4666. This, together with the
presence of a flat ($\G \sim 1.3$) continuum in the same nuclear region that
emits the line, suggests the existence of a strongly absorbed (Compton-thick)
AGN, whose intrinsic 2-10 keV luminosity is $L_{2-10} \magcir 2 \times 10^{41
}$ erg s$^{-1}$.

Compared with other composite AGN/SB galaxies, NGC~4666 shows an important
difference. For example, in NGC~6240 and Arp~299 the AGNs have high intrinsic X-ray
luminosities ($\sim 10^{43}-10^{44}$ erg s$^{-1}$) and are seen according to the
type-1 view (Vignati et al. 1999; Della Ceca et al. 2002). Therefore in these
galaxies the AGN dominates the emission at $E \magcir 2$ keV: the highly-absorbed
primary continuum, transmitted through an obscuring cold medium, emerges from the
absorber and dominates the emission at $E \magcir 10$ keV; and the unabsorbed
primary continuum, scattered into the line-of-sight by a warm gas located outside
the absorbing medium, dominates at $2\, {\rm keV} \mincir E \mincir 10\, {\rm keV}$.
Only at $E \sim 1$ keV can the SB be recognized from the thermal signatures
of the wind components. (A similar situation is encountered in composite AGN/SB-powered
ULIRGs: Franceschini et al. 2003 and Persic et al. 2004). The situation in NGC~4666
is very different. Here the AGN has a much lower (estimated) intrinsic luminosity
($\sim$2$ \times 10^{41}$ erg s$^{-1}$) and is seen in type-2 orientation (the
galaxy is edge-on). The combination of these two factors makes the AGN contribution 
to the 2-10 keV emission of NGC~4666 quite small, $\sim$15$\%$, so the SB -- 
far from being outshone by the AGN -- dominates the integrated 0.3-10 keV continuum. 

The case that a low-luminosity type-2 AGN is identified in the middle of the 
overwhelming glare of a SB makes NGC~4666 an observationally interesting 
and quite rare case. This occurrence owes to the simultaneous use of satellite 
data having widely different characteristics. The low-resolution {\it BeppoSAX} data have 
revealed the dominant continuum component integrated over the disk, therefore 
giving information on the disk-wide SB; they have also revealed the existence 
of a prominent 6.4 keV line, apparently unrelated to the observed continuum. The 
higher-resolution {\it XMM-Newton} data have shown that the 6.40 keV Fe-$K\a$ line originates 
from the central region and, upon zooming onto such region, they have revealed a 
locally flat continuum, hence uncovering the existence of a Compton-thick type-2 AGN; 
they have further revealed a number of possible X-ray point sources that, intepreted 
as accretion-powered binary systems associated with active ongoing star formation, 
can naturally explain the integrated continuum observed by {\it BeppoSAX}.

\begin{acknowledgement}

We have analyzed observations obtained with {\it XMM-Newton}, an ESA science mission 
with instruments and contributions directly funded by ESA Member States 
and NASA (USA). 
This research has made use of the NASA/IPAC Extragalactic Database (NED) 
which is operated by the Jet Propulsion Laboratory, California Institute 
of Technology, under contract with NASA. 
We acknowledge the {\it BeppoSAX} SDC team for providing pre-processed event files. 
We thank an anonymous referee for his/her suggestions that led to a very 
substantial improvement of the paper. 

\end{acknowledgement}

\bigskip

\def\ref{\par\noindent\hangindent 10pt} 
 
\noindent 
{\bf References} 
\vglue 0.2truecm 

\ref{\small Antonucci R.R.J. 1993, ARA\&A, 31, 473}
\ref{\small Antonucci R.R.J. \& Miller J.S. 1985, ApJ, 297, 621}
\ref{\small Ballo L., Braito V., Della Ceca R., et al. 2004, ApJ, 600, 634}
\ref{\small Boella G., Chiappetti L., Conti G., et al. 1997, A\&AS, 122, 327 }
\ref{\small Colbert E.J.M., Heckman T.M., Ptak A.F., \& Strickland D.K. 2003, 
	ApJ, in press (astro-ph/0305476 v2)}
\ref{\small Dahlem M., Petr M., Lehnert M.D., et al. 1997, A\&A, 320, 731}
\ref{\small Dahlem M., Weaver K.A., \& Heckman T.M. 1998, ApJS, 118, 401}
\ref{\small de Vaucouleurs G., de Vaucouleurs A., Corwin H.G. Jr., et al. 1991, 
	The Third Reference Catalogue of Bright Galaxies, Univ. Texas Press, Austin (RC3)}
\ref{\small Della Ceca R., Ballo L., Tavecchio F., et al. 2002, ApJ, 581, L9}
\ref{\small Dickey J.M., \& Lockman F.J. 1990, ARAA, 28, 215}
\ref{\small Fabbiano G., \& White N.E. 2003, in "Compact Stellar X-Ray Sources", eds.
        W. Lewin \& M. van der Klis (Cambridge University Press) (astro-ph/0307077)}
\ref{\small Fabian A.C., Barcons X., Almaini O., \& Iwasawa K. 1998, MNRAS, 297, L11}
\ref{\small Fiore F., Guainazzi M., Grandi P. 1999, "Handbook for NFI Spectral Analysis" available at: \\   
	ftp://sax.sdc.asi.it/pub/sax/doc/software\_docs/saxabc\_v1.2.ps.gz}
\ref{\small Foschini L., Di Cocco G., Ho, L.C., et al. 2002, A\&A, 392, 817}
\ref{\small Franceschini A., Braito V., Persic M., et al. 2003, MNRAS, 343, 1181}
\ref{\small Frontera F., Costa E., Dal Fiume D., et al. 1997, A\&AS, 122, 357}
\ref{\small Garc{\'\i}a, A.M. 1993, A\&AS, 100, 47} 
\ref{\small George I.M. \& Fabian A.C. 1991, MNRAS, 249, 352}
\ref{\small Gonzales Delgado R.M., Heckman T., \& Leitherer C. 2001, ApJ, 546, 845}
\ref{\small Gonzales Delgado R.M., Heckman T., \& Leitherer C. et al. 1998, ApJ, 505, 174}
\ref{\small Griffiths R.E., Ptak A., Feigelson E.D., et al. 2000, Science, 290, 1325}
\ref{\small Guainazzi M., Matt G., Brandt W.N., et al. 2000, A\&A, 356, 463}
\ref{\small Helou G., Soifer B.T., Rowan-Robinson M. 1985, ApJ, 298, L7}
\ref{\small Iwasawa K., Fabian A.C., \& Matt G. 1997, MNRAS, 289, 443}
\ref{\small Kennicutt R.C. 1998, ApJ, 498, 541}
\ref{\small Kilgard R.E., Kaaret P., Krauss M.I., et al. 2002, ApJ, 573, 138}
\ref{\small Krolik J.H. \& Kallman T.R. 1987, ApJ, 320, L5}
\ref{\small Levenson N.A., Krolik J.H., Zycki P.T., et al. 2002, ApJ, 573, L81}
\ref{\small Liu J.-F., Bregman J.N., \& Seitzer P. 2002, ApJ, 580, L31}
\ref{\small Matt, G., Brandt, W.N., \& Fabian, A.C. 1996, MNRAS, 280, 823}
\ref{\small Matt, G., Guainazzi M., Frontera F., et al. 1997, A\&A, 325, L13}
\ref{\small Parmar A.N., Martin D.D.E., Bavdaz M., et al. 1997, A\&AS, 122, 309}
\ref{\small Persic M., Mariani S., Cappi M., et al. 1998, A\&A, 339, L33}
\ref{\small Persic M. \& Rephaeli Y. 2002, A\&A, 382, 843}
\ref{\small Persic M. \& Rephaeli Y. 2004, in Proc. 5th INTEGRAL Workshop, in press
            (astro-ph/0403548)}
\ref{\small Persic M., Rephaeli Y., Braito V., et al. 2004, A\&A, 419, 849}
\ref{\small Pietsch W., Roberts T.P., Sako M., et al. 2001, A\&A, 365, L174}
\ref{\small Rephaeli Y., Gruber D., Persic M., \& McDonald D. 1991, ApJ, 380, L59}
\ref{\small Roberts T.P., Goad M.R., Ward M.J., et al. 2002b, in "New Visions of
                the X-ray Universe in the XMM-Newton and Chandra Era" (astro-ph/0202017)}
\ref{\small Shapiro S.L., Lightman A.P., \& Eardley D.M. 1976, ApJ, 204, 187}
\ref{\small Schaaf R., Pietsch W., Biermann P.L., et al. 1989, ApJ, 336, 777}
\ref{\small Strickland D.K. \& Stevens I.R. 2000, MNRAS, 314, 511}
\ref{\small Str\"uder L., Briel U., Dennerl K., et al. 2001, A\&A, 365, L18}
\ref{\small Suchkov A.A., Balsara D.S., Heckman T.M., \& Leitherer C. 1996, ApJ, 430, 511}
\ref{\small Sukumar S., Velusamy T., \& Klein U. 1988, MNRAS, 231, 765}
\ref{\small Sunyaev R.A. \& Titarchuk L.G. 1980, A\&A, 86, 121}
\ref{\small Terashima Y. \& Wilson A.S. 2003, ApJ, submitted (astro-ph/0305563)}
\ref{\small Tomisaka K., Habe A., \& Ikeuchi S. 1981, Ap\&SS 78, 273}
\ref{\small Turner M.J.L., Abbey A., Arnaud M., et al. 2001, A\&A, 365, L27}
\ref{\small Turner T.J., Perola G.C., Fiore F., et al. 2000, ApJ, 531, 245}
\ref{\small Urry C.M. \& Padovani P. 1995, PASP, 107, 803}
\ref{\small Veilleux S. 2001, in 'Starburst Galaxies: Near and Far', ed. L.Tacconi \& 
     D.Lutz (Heidelberg: Springer-Verlag), 88}
\ref{Vignati P., Molendi S., Matt G., et al. 1999, A\&A, 349, L57}
\ref{Zezas A.L., Fabbiano G., Rots A.H., \& Murray S.S. 2002, ApJ, 577, 710}

\end{document}